\documentclass[pre,aps,amsmath,amssymb,superscriptaddress,notitlepage,twocolumn]{revtex4-2}
\usepackage[colorlinks,bookmarks=false,citecolor=blue,linkcolor=blue,urlcolor=blue]{hyperref}
\usepackage{epsfig}
\usepackage{graphicx}
\usepackage{bm}
\usepackage{amsmath}
\usepackage{physics}
\usepackage{xcolor}
\usepackage{amsmath, nccmath, mathtools}
\usepackage{hyperref}

\usepackage{float}
\usepackage{lipsum}
\usepackage{caption}
\begin{document} 

\title{Probing the localization effects in Krylov basis}

\author{J. Bharathi Kannan}
\email{bharathikannan1130@gmail.com}
\affiliation{Department of Physics, Indian Institute of Science Education and Research, Pune 411008, India}

\author{Sreeram PG}
\affiliation{Department of Physics, Indian Institute of Science Education and Research, Pune 411008, India}

\author{Sanku Paul}
\affiliation{Department of Physics and Complex Systems, S.N. Bose National Centre for Basic Sciences, Kolkata 700106, India}

\author{S. Harshini Tekur}
\affiliation{Department of Physics, Indian Institute of Science Education and Research, Pune 411008, India}

\author{M. S. Santhanam}
\email{santh@iiserpune.ac.in}
\affiliation{Department of Physics, Indian Institute of Science Education and Research, Pune 411008, India}

\date{\today}

\begin{abstract}
Krylov complexity (K-complexity) is a measure of quantum state complexity that minimizes wavefunction spreading across all the possible bases. It serves as a key indicator of operator growth and quantum chaos. In this work, K-complexity and Arnoldi coefficients are applied to probe a variety of localization phenomena in the quantum kicked rotor system. We analyze four distinct localization scenarios -- ranging from compact localization effect arising from quantum anti-resonance to a weaker form of power-law localization -- each one exhibiting distinct K-complexity signatures and Arnoldi coefficient variations. In general, K-complexity not only indicates the degree of localization, but surprisingly also of the nature of localization. In particular, the long-time behaviour of K-complexity and the wavefunction evolution on Krylov chain can distinguish various types of observed localization in QKR. In particular, the time-averaged K-complexity and scaling of the variance of Arnoldi coefficients with effective Planck's constant can distinguish the localization effects induced by the classical regular phase structures and the dynamical localization arising from quantum interferences. Further, the Arnoldi coefficient is shown to capture the transition from integrability to chaos as well. This work shows how localization dynamics manifests in the Krylov basis.
\end{abstract}
\maketitle

\section{Introduction}

Study of complexity in quantum systems has attracted significant attention in recent years, primarily from two perspectives: quantum many-body physics and quantum computation \cite{bernstein1993quantum, nielsen2005geometric,dowling2006geometry, brandao2021models,rabinovici2021operator, rabinovici2022krylov,parker2019universal,susskind2016computational,brown2018second,NielsenChuang,berthiaume2001quantum,hashimoto2023krylov,iaconis2021quantum,haferkamp2022linear}. Several measures have been proposed to quantify complexity, such as circuit complexity \cite{nielsen2005geometric, nielsen2006quantum}, entropy-based measures of quantum states \cite{NielsenChuang}, Quantum Kolmogorov measure \cite{berthiaume2001quantum}, and the focus of this paper -- Krylov complexity (K-complexity) \cite{parker2019universal}. Complexity in quantum systems is often defined as the effort required to transform a simple reference quantum state or operator into a target state or operator using a predefined set of operations \cite{NielsenChuang}.\\

K-complexity was originally developed as a method for efficiently computing exponentials of matrices \cite{hochbruck1997krylov} but has since become a powerful tool for studying operator growth in quantum systems \cite{rabinovici2022krylov, barbon2019evolution, bhattacharya2023krylov, hornedal2022ultimate, erdmenger2023universal, barbon2019evolution, kim2022operator, fan2022universal, ballar2022krylov}.  It quantifies the evolution of an operator by analyzing its spread over a set of vectors called the Krylov basis (K-basis). Unlike measures like circuit complexity, which require arbitrary tolerance parameters for boundedness, K-complexity is parameter-free \cite{ali2020chaos}. In maximally chaotic systems, K-complexity is shown to have an exponential growth up to the scrambling time, mirroring the behaviour of out-of-time-order correlators \cite{parker2019universal, maldacena2016bound}. For non-integrable systems, this growth transitions to linear behaviour and saturates beyond the Heisenberg time \cite{barbon2019evolution}. Its connection to circuit complexity has been explored in \cite{craps2024relation,lv2024building}, where it measures operator growth in the Heisenberg picture via successive nested commutators of the system Hamiltonian with an initially localized observable. Additionally, it serves as a key diagnostic for chaotic dynamics in quantum many-body systems without classical analogues.\\

Expanding on this notion of operator complexity, the concept of quantum state complexity called spread complexity was introduced \cite{balasubramanian2022quantum}.  The evolution of the spread complexity over time is a metric for how quickly a chosen reference state spreads under its dynamics within the K-basis. The K-complexity, defined initially for operators, can also be understood as a state complexity on a doubled Hilbert space once the ambiguity over the choice of the inner product is eliminated \cite{balasubramanian2022quantum}. For consistency, we refer to spread complexity simply as K-complexity in this work. \\

While the K-basis construction using the iterative methods works well for time-independent Hamiltonians, it fails for time-dependent systems. However, for Floquet systems, due to their periodic drive, a straightforward generalization can be made, where the unitary is the generator of the dynamics instead of the Hamiltonian \cite{yates2021strong}. Recently, this method was used to study integrable to chaos transition in the quantum kicked rotor (QKR) model \cite{nizami2023krylov}. QKR is a prototypical model to investigate quantum chaos and dynamical localization \cite{santhanam2022quantum}. Unlike the classical kicked rotor, which exhibits unbounded diffusive energy growth due to classical chaos, the quantum version excites only a finite number of momentum states, and the momentum distribution follows a characteristic exponentially decaying profile. This phenomenon is called dynamical localization \cite{santhanam2022quantum, casati2005stochastic}, and it arises purely from quantum coherence and, as a result, is highly sensitive to the effects of decoherence \cite{maurya2024asymmetric, maurya2022interplay, sarkar2017nonexponential,paul2019decoherence,paul2020CKR}. Quantum dynamical localization can be interpreted as a disorder-free analog of Anderson localization \cite{grempel1984quantum} and has also been experimentally observed \cite{moore1994observation}. \\

In this paper, we examine the QKR model with cylindrical boundary conditions in momentum space, utilizing two distinct potentials: (i) a smooth cosine potential corresponding to the canonical QKR (CQKR), the most widely studied version, and (ii) a non-smooth potential known to exhibit power-law localized states, referred to as the singular QKR (SQKR). Through these two models, we provide a comprehensive analysis of four different types of localizations: (a) classical-induced localization (CIL), (b) dynamical localization (DL), (c) quantum anti-resonance (AR), and (d) power-law localization (PL). The only prior study on K-complexity in the QKR model, conducted with toral boundary conditions, utilized K-complexity and the variance of Arnoldi coefficients to differentiate between integrable and chaotic regimes \cite{nizami2023krylov}. K-complexity depends solely on the Hamiltonian and the initial state, mirroring the key factors that govern classical chaos: system parameters (in H) and phase space initial conditions. While this suggests a potential role for K-complexity as a chaos diagnostic, it does not conclusively establish it. In contrast, other measures, such as the Loschmidt echo, are sensitive to specific Hamiltonian perturbations, while circuit complexity involves ambiguities like tolerance thresholds, the choice of elementary gates, and the metric on the operator manifold. In this work, we focus on investigating how different localization effects manifest in the K-basis and how these effects influence quantities such as K-complexity and Arnoldi coefficients across the diverse dynamical regimes exhibited by these two systems.\\

The paper is organized as follows: Section \ref{Sec:2} provides an overview of K-complexity and its construction for Floquet systems. Section \ref{Sec:3} introduces the studied models and explores different types of localization in the momentum basis, while Section \ref{Sec:4} examines localization in the K- basis. In Section \ref{Sec:5}, we analyze the time evolution of K-complexity and Arnoldi coefficients across different dynamical regimes. Section \ref{Sec:6} investigates the long-time behaviour of these complexity measures and their scaling with system parameters. Finally, Section \ref{Sec:7} summarizes the key results and outlines future directions.

\section{Construction of K-basis for floquet systems}
\label{Sec:2}

In this section, we briefly outline the concepts of K-basis construction for floquet systems using Arnoldi iteration, along with K-complexity and Arnoldi coefficients, which are central to our analysis \cite{nizami2023krylov, nizami2024spread}. K-complexity quantifies the complexity of operators \cite{parker2019universal} or states \cite{balasubramanian2022quantum} based on their spread in the K-basis. In this work, we focus solely on state complexity, though the extension to operator complexity is straightforward. Generally, complexity measures are inherently basis-dependent and not unique \cite{balasubramanian2022quantum}; however, the K-basis holds a special significance as it minimizes the spread across all possible bases \cite{balasubramanian2022quantum}. It has been shown that within a given time interval $[0, T]$, K-complexity cannot be reduced by adopting any alternative basis \cite{balasubramanian2022quantum}. In time-dependent systems, the evolution operator is expressed as Dyson's time-ordered exponential, $U(t) = \mathcal{T} \exp\left(-i \int_0^t H(t') \, dt'\right)$, unlike the simple exponential used in time-independent systems, where \(\mathcal{T}\) denotes the time-ordering operator. For periodically driven systems, the Floquet formalism simplifies this by having an effective time-independent Floquet Hamiltonian $H_F$ such that the evolution operator over one period can be expressed as $U_{F}(T) = \exp(-i H_F T)$. This enables efficient construction of the K-basis by avoiding the need for explicit time-ordering and leveraging the periodicity to analyze the dynamics. This approach, introduced in Ref. \cite{yates2021strong}, extends the Lanczos algorithm to Floquet systems by employing Arnoldi iteration to construct the K-basis. \\

Since we focus on a kicked model, namely QKR, K-complexity can be studied by analyzing the dynamics at stroboscopic times and time evolution is dictated through a floquet matrix $U_F$. Let the normalized initial state be $\ket{\psi_0}$, time evolution is obtained by stroboscopically probing the system at equal time intervals, forming the sequence $\{\ket{\psi_0}, U_F\ket{\psi_0}, U_F^2\ket{\psi_0}, \dots\}$, where the $n$-th element represents the state after $n$ discrete time steps. The Arnoldi iteration technique \cite{arnoldi1951principle} can be employed to construct an orthonormal basis from this set which generalises the Lanczos procedure beyond the Hermitian case. This method applies a Gram-Schmidt-like orthogonalization process, explicitly orthogonalizing each vector against all previous basis vectors. Then, the $n$-th orthonormal K-basis vector is given by 
\begin{equation}
    \ket{K_n} = \frac{1}{h_{n,n-1}}\left( U_{F} \ket{K_{n-1}} - \sum_{i=0}^{n-1} h_{i,n-1}\ket{K_i} \right); n\geq 2,
\end{equation}
where $h_{j,k} = \bra{K_j} U_F \ket{K_k}$ are the Arnoldi coefficients, analogous to the Lanczos coefficients in the time-independent case. The normalization constants $h_{n,n-1}$ represent the norm of the $n$-th Krylov vector and are very similar to the Lanczos $b_n$'s. This generalization enables the construction of orthonormal vectors using non-Hermitian operators and has been applied to Floquet systems \cite{nizami2023krylov, yates2021strong, nizami2024spread}, and to open quantum systems \cite{bhattacharya2022operator, bhattacharjee2023operator}. \\

The state at time $t$ is $\ket{\psi(t)} = U_F^t \ket{\psi_0}$. The K-complexity of  $\ket{\psi(t)}$ is quantified by its overlap with the K-basis vectors, expressed as
\begin{equation}
\mathcal{C}(t) = \sum_{n=0}^{D_k-1} n \left| \bra{K_n} \ket{\psi(t)} \right|^2 = \sum_{n=0}^{D_k-1} n \left| \phi_n(t) \right|^2.
\end{equation}
In this, $D_k$ is the number of K-basis vectors generated using $U_F$, and 
\begin{equation}
\phi_n(t) = \bra{K_n} \psi(t) \rangle
\label{eq:phin}
\end{equation}
represents the projection of $\phi_n(t)$ on the $n$-th K-basis vector such that $\sum_n |\phi_n(t)|^2 = 1$. The temporal spread of the state over the K-basis at time $t$ is quantified by $|\phi_n(t)|^2$. In a later section, we will use the participation ratio of  $|\phi_n(t)|^2$ to quantify the degree of localization. \\

In the Krylov construction, the dynamics of a quantum system is mapped to a particle-hopping tight-binding model on a one-dimensional lattice, with the Lanczos coefficients representing the hopping amplitudes for time-independent Hamiltonians \cite{parker2019universal}. This idea can be extended to Floquet systems \cite{nizami2023krylov}. Using this framework, it has been shown that the K-complexity growth in a Floquet system is at most linear in time. This is due to the fact that the particle on the Krylov chain can only hop one step to the right at each time step.
After $t$ time steps, the particle can be at most $t$ lattice sites away from its initial position, leading to a K-complexity growth that is bounded by $t$ and, thus, at most linear in time. This linear growth in operator complexity has also been demonstrated in maximally chaotic dual-unitary models \cite{suchsland2023krylov} .\\

Another intriguing and less studied aspect is the variance of the Lanczos $b_n$'s \cite{rabinovici2022krylov} and, in our case, the $h_{n,n-1}$ Arnoldi coefficients, which reflect key signatures of system dynamics. The variance $\sigma^2(h_{n,n-1})$ quantifies the magnitude of fluctuations in $h_{n,n-1}$ and is defined as
\begin{equation} 
\sigma^2(h_{n,n-1}) = \frac{1}{D_k} \sum_{n=0}^{D_k-1} \left(h_{n,n-1} - \overline{h_{n,n-1}}\right)^2
\end{equation}
where $\overline{h_{n,n-1}}$ denotes the mean value of $h_{n,n-1}$. Through the mapping to the Krylov chain, a smaller variance implies less erratic behaviour of Lanczos coefficients, and this yields a larger localization length, so the Krylov chain is more delocalized. This typically occurs when the system is more chaotic. Notably, Arnoldi coefficients exhibit stronger fluctuations in near-integrable regimes compared to chaotic ones\cite{nizami2023krylov}. This paper utilises K-complexity and Arnoldi coefficients to investigate different types of localization effects in the QKR model. \\

\section{Spectrum of localization in QKR}
\label{Sec:3}
\subsection{QKR models}
In this section, we provide a brief overview of the types of localizations occurring in the kicked rotor model \cite{casati2005stochastic, santhanam2022quantum}. Physically, the model describes a particle subjected to periodic kicks and is closely connected to the Anderson model \cite{santhanam2022quantum}, which describes the electronic transport in a disordered lattice and displays localization of the electronic states. Hamiltonian governing QKR is given by
\begin{equation}
    H(t) = \frac{\hat{p}^2}{2} + V(\hat{x})\sum_{n=0}^\infty \delta(t - nT),
\label{Eq:Ham_QKR}
\end{equation}
where $\hat{x}$ and $\hat{p}$ are position and momentum operators, $V(\hat{x})$ is the kicking potential. This model provides a simple yet rich framework to study quantum chaos and investigate how structures in classical phase space influence the quantum properties of non-integrable systems. In this work, we explore two variants of the kicked rotor model with potentials given as 
\begin{enumerate}
    \item $V(\hat{x}) = K \cos(\hat{x})$, ~~~Canonical QKR (CQKR) \cite{santhanam2022quantum},
    \item $V(\hat{x}) = K |\hat{x}|^\alpha$, ~~~Singular QKR (SQKR) \cite{garcia2005anderson}.
\end{enumerate}
The parameters $K$, $T$  and $\alpha$ represent the kicking strength, kick period of kicking and exponent in the potential, respectively. In both models, we impose cylindrical boundary conditions, where $x \in [0, 2\pi]$ and $p \in (-\infty, \infty)$. Without loss of generality, we set the time period of kicking $T = 1$ throughout this work. CQKR model has a smooth kicking potential and is extensively studied in the context of Anderson localization and the metal-insulator transition in cold atoms experiments \cite{chabe2008experimental, lopez2012experimental}. SQKR is a singular version of the standard kicked rotor and is known to display power-law localization \cite{garcia2005anderson,sanku2018powerlaw}. The dynamics of kicked rotor systems can be analyzed using Floquet theory, leveraging the periodic nature of the time-dependent Hamiltonian. The evolution operator over one period is
\begin{equation}
    U_{F} = \exp({-i\frac{\hat{p}^{2}}{2\hbar_{s}}}) ~ \exp({-i\frac{V({\hat{x}})}{\hbar_{s}}}),
    \label{eqn:floquet_op}
\end{equation}
where the first exponential term describes free evolution and the second accounts for the kicks. The scaled Planck's constant $\hbar_{s}$ can be tuned to transition from quantum to classical limit. For CQKR, the Floquet operator $\langle m_2|U_{F}|m_1 \rangle$ in the basis of the momentum eigenstates ($|m\rangle = e^{imx} / \sqrt 2\pi$) is
\begin{equation}
    \widehat{U}_F(m_1, m_2)=\exp \left(-\frac{i}{2} \hbar_s m_2^2\right) i^{m_2-m_1} J_{m_2-m_1}\left(\frac{K}{\hbar_s}\right) .
    \label{Eq:Floq_CQKR}
\end{equation}
Note that $\widehat{U}_F$ is a diagonally dominant matrix since $J_{m_2-m_1}(.) \rightarrow 0$ for $\left|m_2-m_1\right| \gg 1$. This property makes it an effectively banded matrix. The eigenvectors of such a matrix can be expected to get significant contributions only from a narrow band of basis states, implying localized excitations \cite{santhanam2022quantum}. In the case of the SQKR, the Floquet operator does not have an analytical form in the momentum basis under cylindrical boundary conditions. Therefore, it is constructed numerically using the split-operator technique by acting on the momentum basis vectors with the Floquet operator given in Eq.(\ref{eqn:floquet_op}) with the singular potential \cite{santhanam2022quantum, sanku2018powerlaw}. The K-basis is constructed using the unitaries in depending on the specific type of localization we aim to investigate. \\

\begin{figure}[t]
    \centering
    \includegraphics[width=\linewidth]{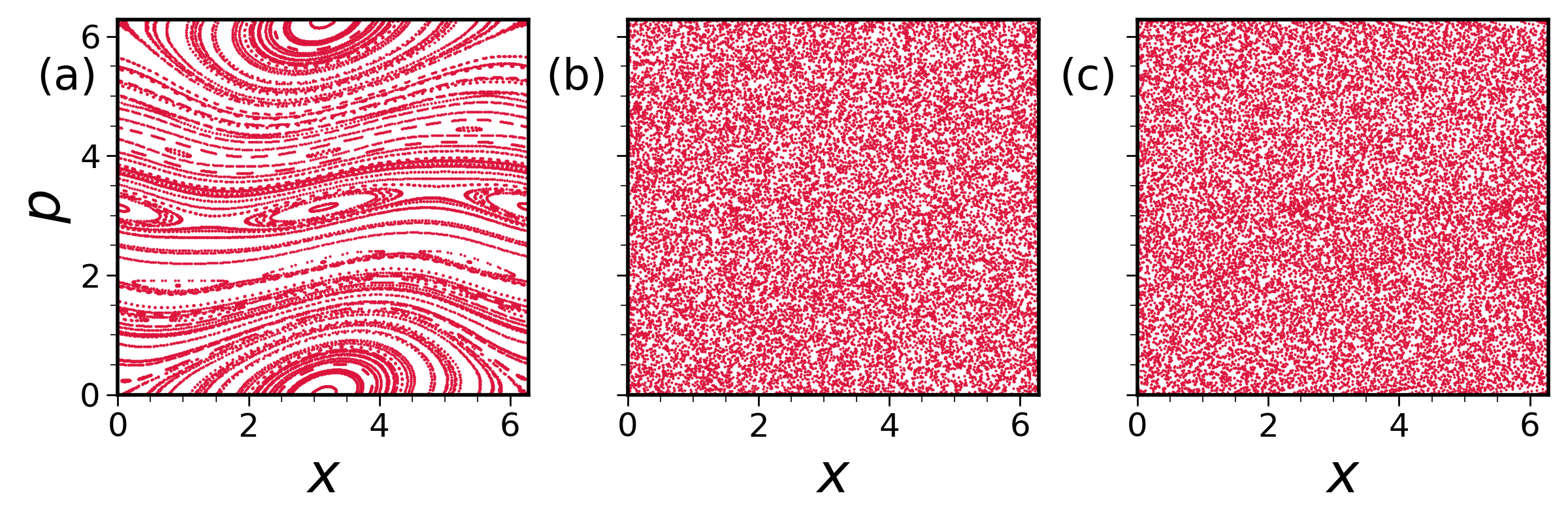}
    \includegraphics[width=\linewidth]{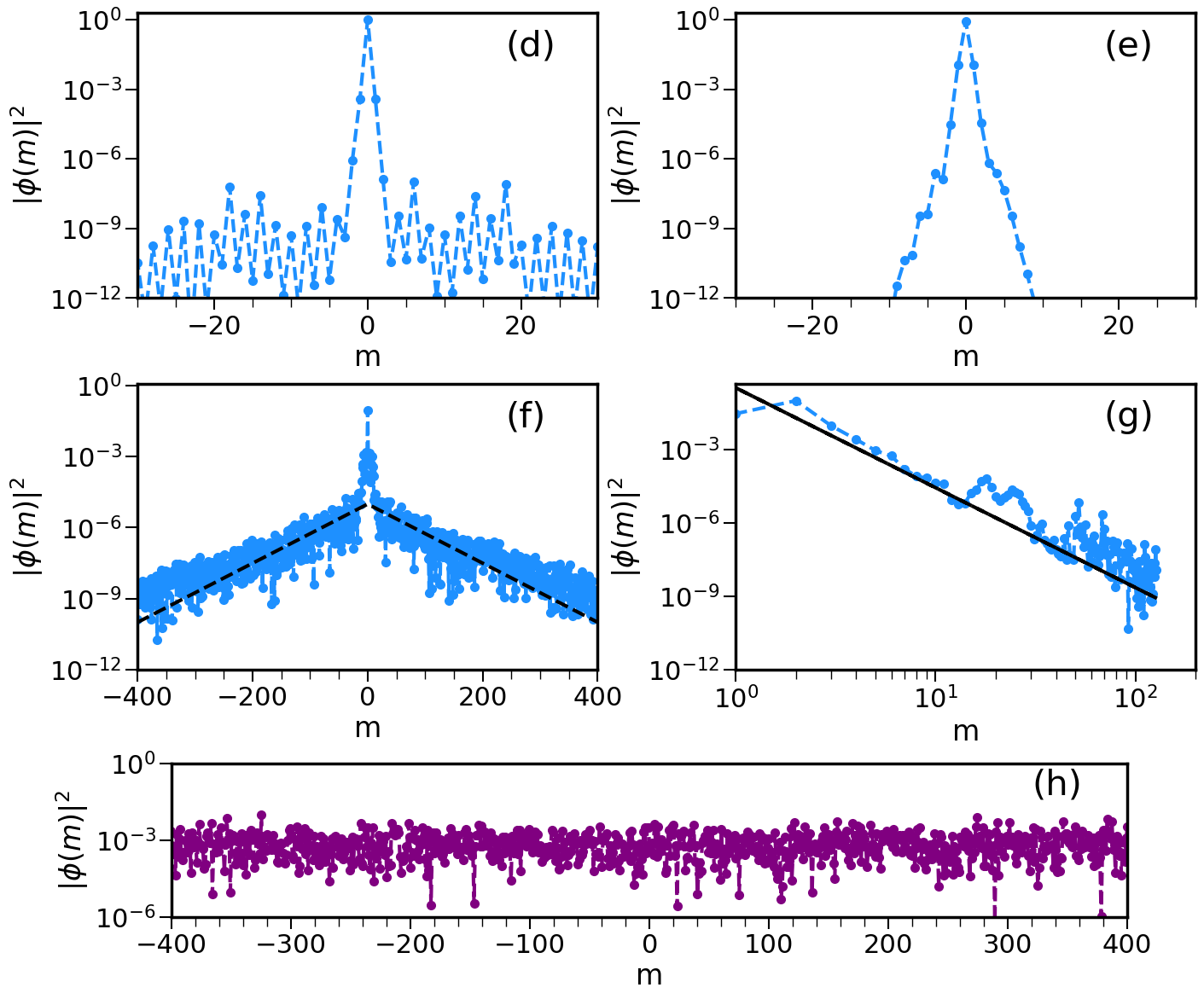}
    \caption{Top row: Phase space plots of the standard kicked rotor and singular kicked rotor models. The x-axis represents the angular position $x$, while the y-axis shows the conjugate momentum $p$. 
    (a) KR with $K = 0.5$, showing regular motion around primary islands,
    (b) KR with $K = 8$, exhibiting chaotic behaviour with large diffusion in phase space,
    (c) Singular KR with $K = 1$, $\alpha = 0.5$, demonstrating complete chaos in phase space.
    Bottom rows ((d)-(g)): Probability density $|\phi(m)|^2$   of the averaged eigenstates in momentum space for different localization regimes in the Log-scale. $m$ denotes the momentum index.
    (d) Quantum anti-resonance (AR) for $K = 0.5$, $\hbar_s = 2\pi$,
    (e) Classical-induced localization (CIL) for $K = 0.5$, $\hbar_s = 1$, 
    (f) Dynamical localization (DL) for $K = 8$, $\hbar_s = 1$ (dotted line: exponential fit),
    (g) Power-law localization (PL) in the singular kicked rotor (SQKR) model with $K = 1$, $\alpha = 0.5$ (dotted line: power-law fit).
    (h) Delocalized eigenvector, representing a random complex eigenstate with uniform probability distribution in momentum space. }
    \label{fig:loc_momentum}
\end{figure}

\subsection{Variety of localization}
Earlier studies have examined K-complexity in the QKR model, investigating the chaotic and regular regimes under toral boundary conditions \cite{nizami2023krylov}. In this work, we focus on the QKR model with cylindrical boundary conditions to understand the behaviour of K-complexity measures in the regimes characterised by different types and degrees of localization. Key localization effects observed in the QKR model are (i) quantum anti-resonance (AR), (ii) classical-induced localization (CIL), (iii) dynamical localization (DL), and (iv) power-law localization (PL). Quantum anti-resonance, with its almost delta-function-like profile, is the strongest form, while power-law localization, with its power-law wavefunction profile, is a relatively weaker form. These offer a comprehensive landscape to probe K-complexity.\\

Before examining localization effects in the K-basis, it is crucial to understand how these effects manifest in the Floquet matrix eigenstates in momentum space. To build this intuition, we analyze the averaged eigenstates obtained by diagonalizing the Floquet matrix in momentum space, revealing the localization type depending on the choice of parameters as summarized in Table \ref{tab:localization_summary}. The parameters are chosen to represent each regime. The top panel of Fig. (\ref{fig:loc_momentum}) shows the classical phase space for CQKR: (a) for $K=0.5$ and (b) for $K=8$. Note that $K=0.5$ falls in the regular regime, while $K=8$ is dominated by chaos. Panel (c) depicts the phase space of SQKR with $K=1$ and $\alpha=0.5$, and is pre-dominantly chaotic. 
Fig \ref{fig:loc_momentum}(d-g) shows $|\phi(m)|^2$ in momentum space for the parameters in Table \ref{tab:localization_summary}.

\begin{table}[h!]
\centering
\caption{Summary of the different localization regimes, their parameters, and the corresponding models used in Fig.\ref{fig:loc_momentum}. The parameters for each localization type remain fixed throughout the paper unless stated otherwise.}
\begin{tabular}{|p{4.2cm}|p{3.1cm}|p{1cm}|}
\hline
\textbf{Regime} & \textbf{Parameters} & \textbf{Model} \\
\hline
Anti-Resonance (AR) & $(K,\hbar_s)=(0.5,2\pi)$ & CQKR \\
\hline
Classical Induced Localization (CIL) & $(K,\hbar_s)=(0.5,1)$ & CQKR \\
\hline
Dynamical Localization (DL) & $(K,\hbar_s)=(8,1)$ & CQKR \\
\hline
Power Law Localization (PL) & $(K, \alpha, \hbar_s) = (1, 0.5, 1)$ & SQKR \\
\hline
\end{tabular}
\label{tab:localization_summary}
\end{table}

{\it Quantum Anti-resonance} (AR) is a purely quantum phenomenon and arises when $\hbar_s = (4n + 2)\pi$, where $n \in \mathbb{Z}$. When this condition is met, the free evolution part of $\widehat{U}_F$ in Eq. \ref{Eq:Floq_CQKR} oscillates between +1 and -1 for even and odd momentum sites \cite{santhanam2022quantum}. Then, the eigenstates of the floquet matrix are sharply localized in momentum basis, irrespective of the nature of corresponding classical dynamics. This is shown in Fig. \ref{fig:loc_momentum}(d).

{\it Classical-induced localization} is a semiclassical effect that usually arises for small kick strengths $K < 1$ due to the influence of regular classical phase space structures on the evolving quantum dynamics. In this case, the phase space structures ''trap'' the wave function, effectively suppressing it from spreading over all the momentum states. This manifests as a semiclassical localization and is shown in Fig. \ref{fig:loc_momentum}(e).

{\it Dynamical localization} is an emergent form of quantum interference effect in momentum space.  It is a form of Anderson-type localization and manifests with an exponential profile $\sim \exp(-n/n_l)$, where $n_l$ is the localization length. DL manifests for $t > t_b$, where $t_b$ is a characteristic timescale known as the break time, marking the onset of pronounced quantum effects \cite{santhanam2022quantum}. This is usually observed in the classically chaotic regime with $K>1$ and has been experimentally realized as well. This is shown in Fig. \ref{fig:loc_momentum}(f).

{\it Power-law localization} is observed in the SQKR model (and not in the CQKR), driven by quantum effects and singularity in the potential. Though quantum in origin, this is a weaker form of localization effect since the wavefunction displays a power-law profile $\sim n^{-\gamma}$ over the momentum basis states. This is shown in Fig. \ref{fig:loc_momentum}(g). For comparison, Fig. \ref{fig:loc_momentum}(h) presents a delocalized eigenvector, representing a random complex eigenstate with a uniform probability distribution in momentum space, clearly contrasting the different types of localization with delocalization.\\

One signature of quantum chaos is through the statistical distribution of adjacent energy level spacings described by the Gaussian ensembles of random matrix theory. In contrast, for integrable quantum systems, spacings are Poisson distributed. Except for the AR case (which has only two levels), all other types of localization exhibit a level spacing distribution close to Poisson due to localization effects. Consequently, level statistics alone cannot distinguish between the varying degrees of localization \cite{santhanam2022quantum, garcia2005anderson} (see Appendix \ref{app:sprectral_stats}). \\

\begin{figure*}[t]
    \centering
    \includegraphics[width=0.49\linewidth, height=0.16\linewidth]{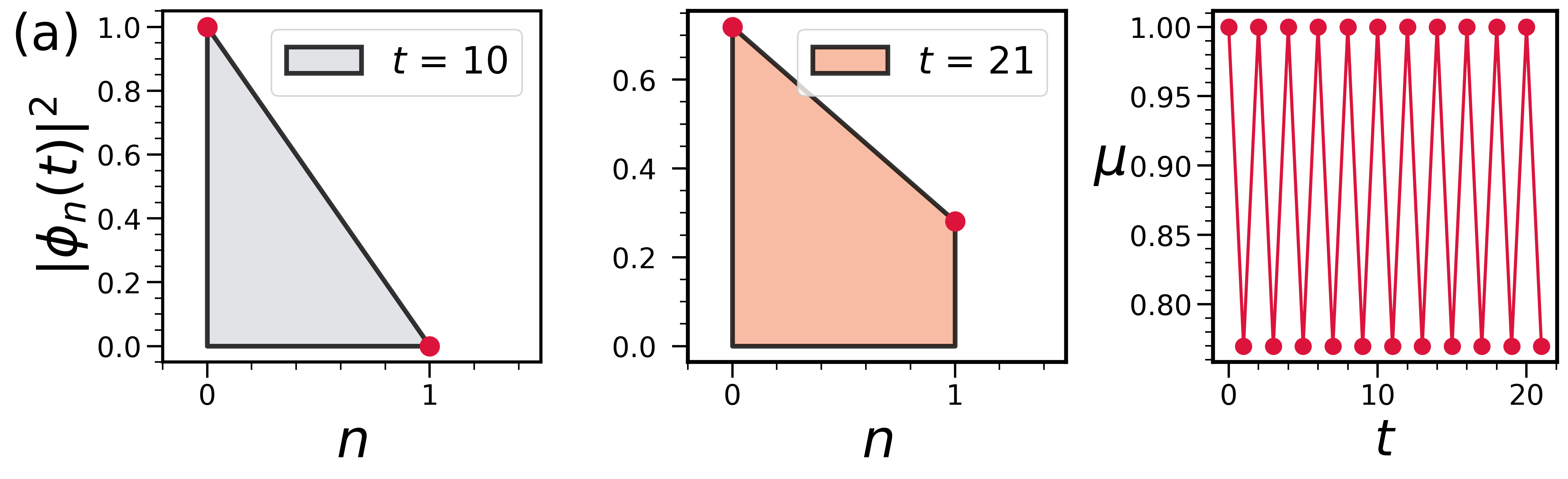}
    \includegraphics[width=0.49\linewidth, height=0.16\linewidth]{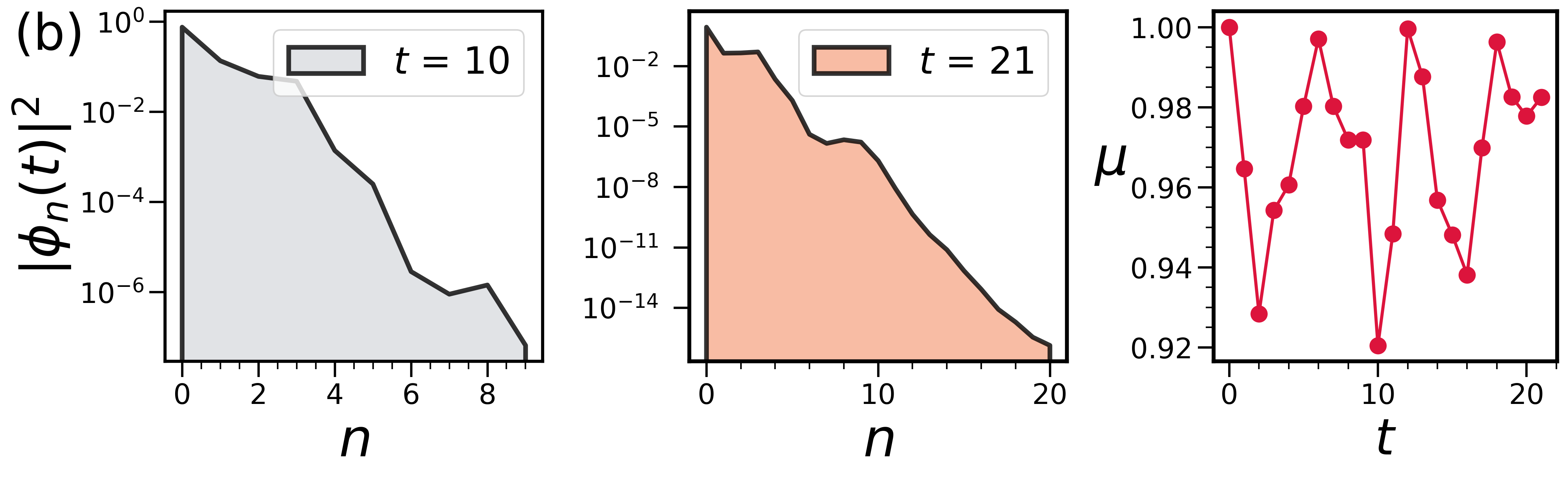}
    \includegraphics[width=0.49\linewidth, height=0.16\linewidth]{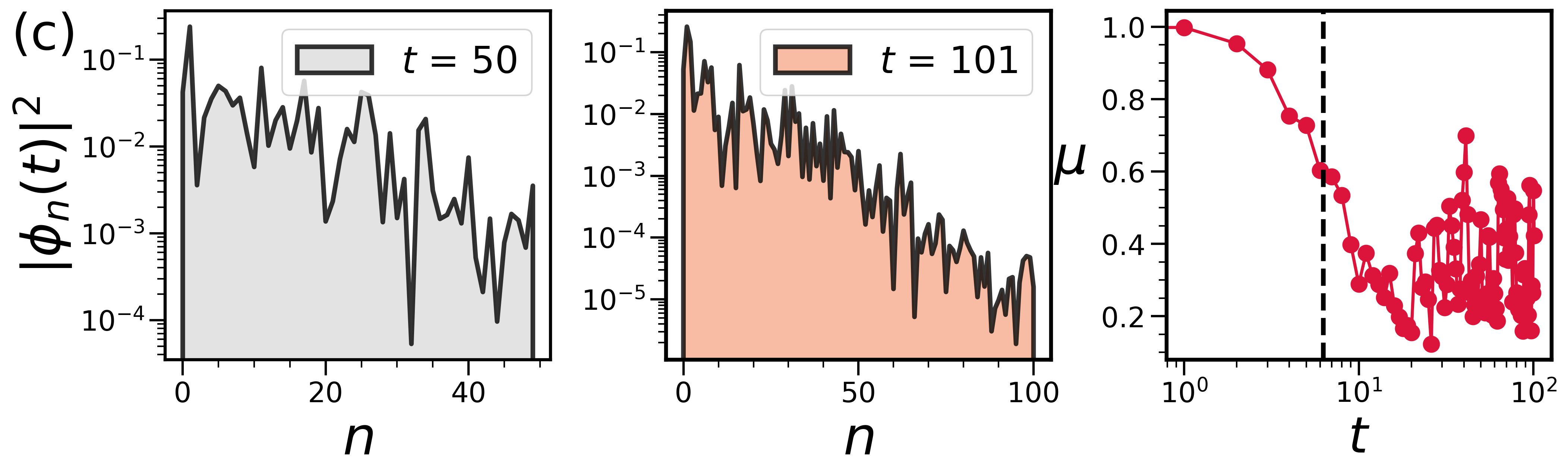}
    \includegraphics[width=0.49\linewidth, height=0.16\linewidth]{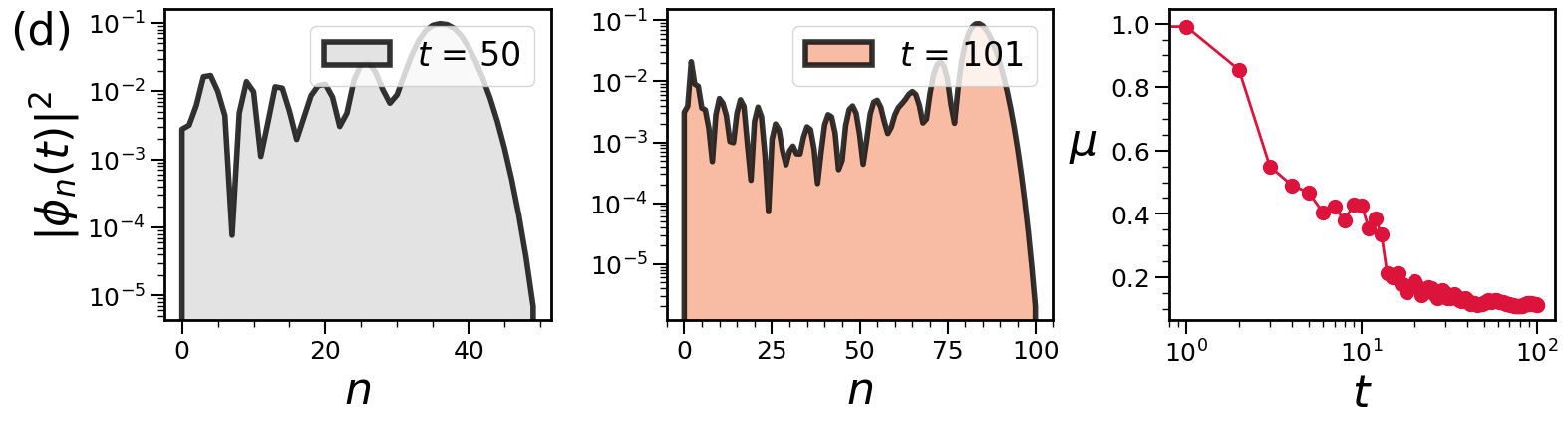}
    \caption{Probability density $|\phi_{n}(t)|^{2}$ and inverse participation ratio (IPR) in the K-basis as a function of the number of kicks ($t$) for different parameter sets. (a) AR case, where red dots highlight strong localization at the first and second K-basis vectors. (b) CIL case, showing relatively weaker localization in the K-basis compared to the AR case. (c) DL case, with the black dotted line marking the break time, beyond which localization sets in. (d) Singular kicked rotor case with $K = 1$, displaying power-law localized states and corresponding IPR, indicating weak localization effects.}
    \label{fig:wf_Kbasis}
\end{figure*}

\section{Wavefunction spread in K-Basis}
\label{Sec:4}
Beyond localized states in momentum space (Fig. \ref{fig:loc_momentum}), we examine the dynamics of the time-evolving wavefunction in the K-basis. Fig \ref{fig:wf_Kbasis} illustrates $|\phi_n(t)|^2 = |\bra{K_n} \ket{\psi(t)}|^2$ in the Krylov chain for the localized phases. Each panel of Fig. \ref{fig:wf_Kbasis} has 3 subplots, in which the left and centre subplots display $|\phi_n(t)|^2$, probability density at site $n$ in K-basis, for two distinct time steps. This analysis sheds light on the evolution of contributions from higher K-basis vectors in the localization regimes. The degree of localization in the K-basis is quantified using the inverse participation ratio (IPR), defined as:

\begin{equation}
    \mu(t) = \mathrm{IPR}(t) = \sum_{n} |\phi_{n}(t)|^4.
\end{equation}
In each panel of Fig \ref{fig:wf_Kbasis}, the right subplot shows $\mu(t)$ to infer how the degree of localization evolves over time.
If $\mu(t) \to 1$, then it indicates strong localization, as wavefunction concentrates over fewer K-basis vectors. A small IPR value of $\mu \to 0$ corresponds to delocalization, as the wave function is spread over many K-basis vectors.\\  

Fig \ref{fig:wf_Kbasis}(a) shows $|\phi_n(t)|^2$ at an even and odd time step for the AR regime. In this case, the system dynamics is captured only by two K-basis vectors. This is highlighted in the rightmost subplot of Fig. \ref{fig:wf_Kbasis}(a), where $\mu(t)$ oscillates between two discrete values $\mu_1$ and $\mu_2$. The separation  $\mu_1 - \mu_2$ depends on the value of $K$; the larger the $K$, the larger the separation.  These oscillations indicate a strong tendency for non-spreading, i.e., localization in the K-basis, and periodicity of two implies that dynamics can be effectively captured using just two K-basis vectors.

In the case of CIL, as seen in Fig \ref{fig:wf_Kbasis}(b), $\ket{\psi(t)}$ evolves slowly in K-basis with significant contribution from $K_0$ (the initial K-basis vector). This suggests that only a few K-basis vectors are sufficient to effectively describe the dynamics. The corresponding IPR fluctuates but remains close to unity (right subplot of Fig \ref{fig:wf_Kbasis}(b)), signifying strong localization. However, unlike the AR regime, there are no perfect revivals, indicating weaker localization than in the AR case.

Fig \ref{fig:wf_Kbasis}(c) depicts the DL regime in K-basis. The initial time evolution leads to a spread in the K-basis, more than for AR and CIL regimes, followed by saturation over long times. The IPR, in Fig \ref{fig:wf_Kbasis}(c), exhibits an initial decay followed by oscillation as a function of time, reflecting the initial spreading followed by localization in the K-basis. Notably, the decay of the IPR persists well beyond the break time $t_b$ (indicated by the dashed black vertical line). This indicates that the wavefunction evolution in the K-basis does not mimic that in the natural momentum basis.  Even after DL sets in, the wavefunction may continue to spread within the K-basis, as it has not fully utilized the basis to describe its dynamics. Notably, this saturation is not attributed to the finiteness of the basis but rather to the emergence of dynamical localization. Even if basis size $N$ is significantly increased, the saturation values remain largely unchanged, as demonstrated in Appendix \ref{app:system_size} .\\

In the case of power-law localization, shown in Fig. \ref{fig:wf_Kbasis}(d), a rapid initial spread of the wavefunction in the K-basis is followed by a quicker decay than in cases of AR, CIL and DL. This behaviour is characteristic of a weaker form of localization. Consistent with this observation, $\mu(t)$ rapidly decays, followed by a saturation, as shown in the right subplot of Fig. \ref{fig:wf_Kbasis}(d), at values approaching those of a uniformly delocalized wavefunction in the K-basis. This fast decay indicates a weaker localization effect as the system reaches a state of near-delocalization on the K-basis. \\

\begin{figure}[h!]
    \centering
    \includegraphics[width=0.49\linewidth]{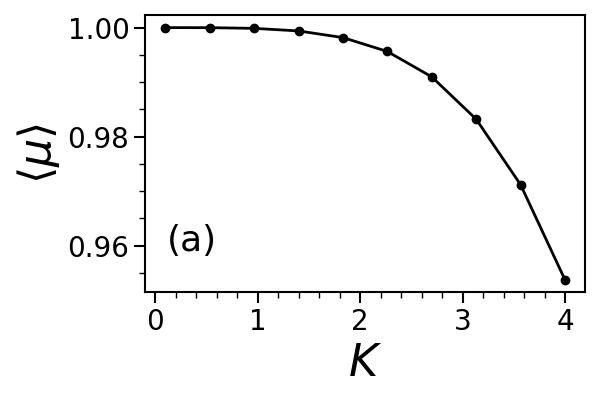}
    \includegraphics[width=0.49\linewidth]{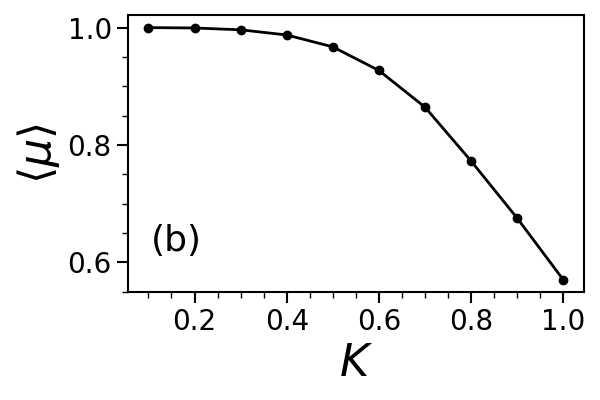}
    \includegraphics[width=0.49\linewidth]{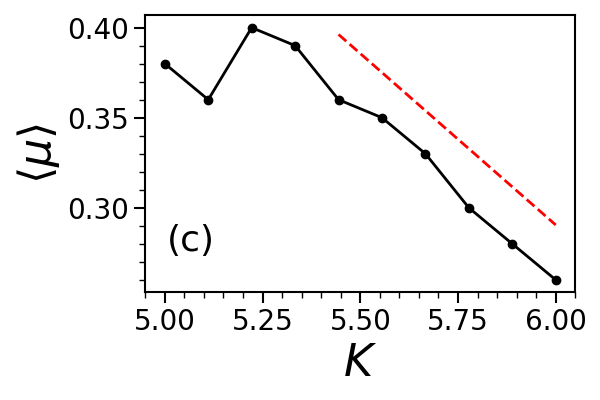}
    \includegraphics[width=0.49\linewidth]{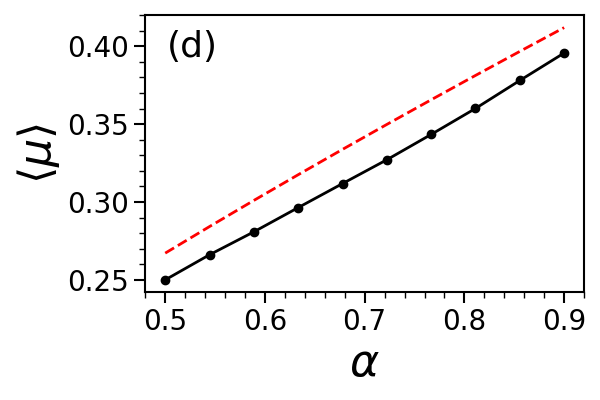}
    \caption{Inverse participation ratio (IPR) in the K-basis as a function of $K$ for different types of localization in CQKR: (a) AR, (b) CIL, and (c) DL. Panel (d) shows the IPR as a function of $\alpha$ for power-law localization (PL) in the SQKR. Higher IPR values indicate stronger localization of the wave function in the K-basis. The red dotted lines in (c) and (d) serve as a guide to highlight the linear trend of the IPR with the respective parameters.}
    \label{fig:IPR}
\end{figure}

How do localization properties change as a function of parameters? Fig \ref{fig:IPR} illustrates IPR as a function of the parameters: $K$ for CQKR and $\alpha$ for SQKR. These parameters govern the degree of localization in their respective localized phases. In Fig. \ref{fig:IPR}(a), $\langle \mu \rangle$ exhibits algebraic decay with $K$, indicating increased overlap with the second K-basis vector as $K$ increases.  A similar trend is observed for CIL in Fig. \ref{fig:IPR}(b), where stronger kicks expand chaotic regions in phase space, enhancing spread in the K-basis. Fig. \ref{fig:IPR}(c) shows a linear decay of $\langle \mu \rangle$ with $K$ on average, due to the fact that the localization length in the momentum basis scales as $K^2$. As $K$ increases, the enhanced spread in the momentum basis translates directly to a greater spread in the K-basis. In the PL regime, the control parameter is $\alpha$. Here, IPR increases with $\alpha$, due to a transition towards stronger localization as seen in Fig \ref{fig:IPR}(d). This trend suggests that $\alpha$ plays a critical role in controlling the degree of power-law localization, with larger $\alpha$ associated with lesser wavefunction spread in the K-basis. \\

\section{Evolution of K-complexity and Arnoldi Coefficients}
\label{Sec:5}

\begin{figure}[h!]
    \centering
    \includegraphics[width=0.48\linewidth]{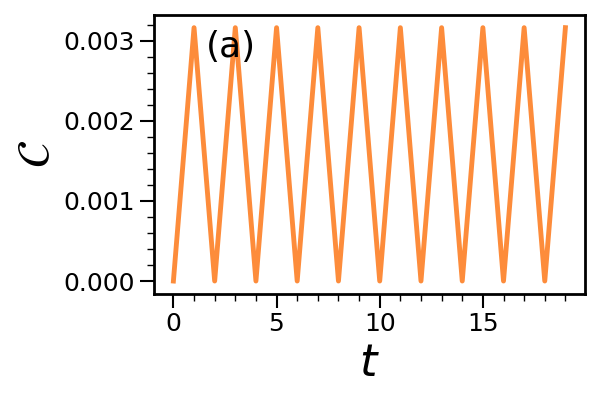}
    \includegraphics[width=0.48\linewidth]{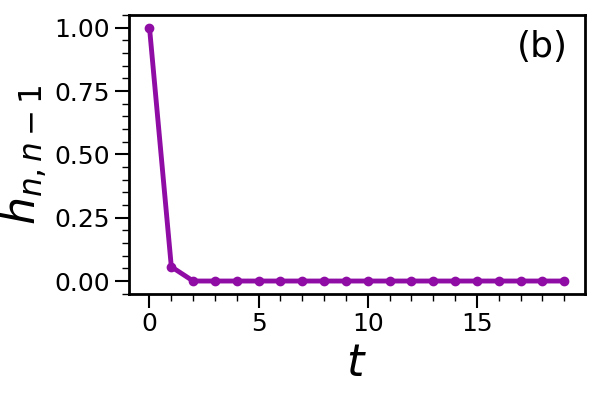}
    \includegraphics[width=0.48\linewidth]{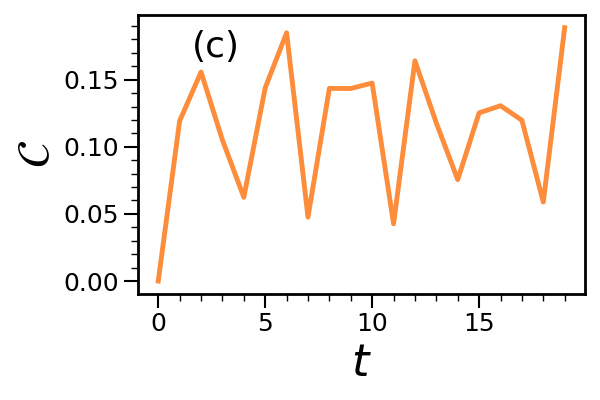}
    \includegraphics[width=0.48\linewidth]{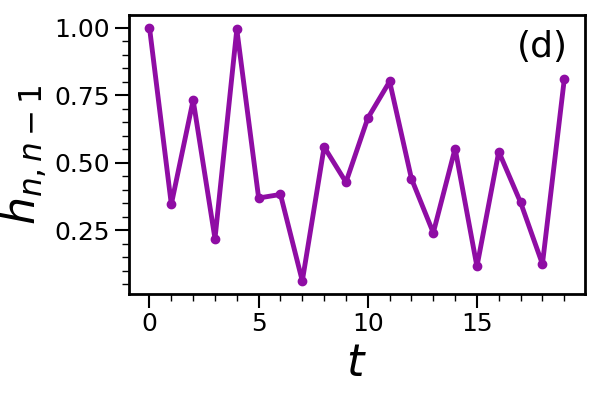}
    \includegraphics[width=0.48\linewidth]{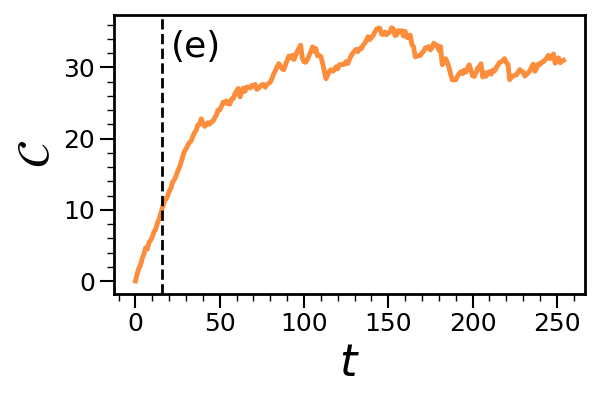}
    \includegraphics[width=0.48\linewidth]{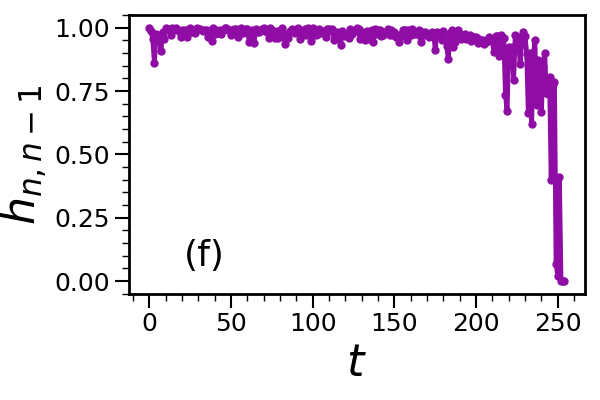}
    \includegraphics[width=0.48\linewidth]{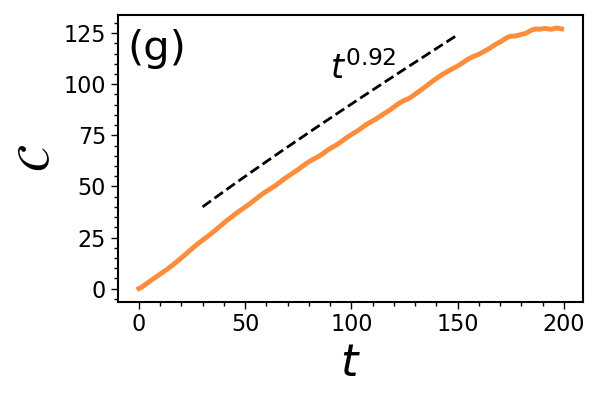}
    \includegraphics[width=0.48\linewidth]{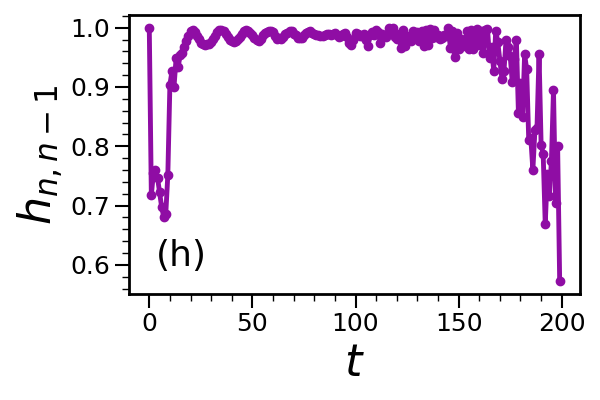}
    \caption{K-complexity ($\mathcal{C}(t)$) and Arnoldi coefficients ($h_{n,n-1}(t)$) across the four regimes of localization: AR, CIL, DL, and PL. The left column shows the time evolution of K-complexity while the right column illustrates the behaviour of Arnoldi coefficients corresponding to each localization regime.}
    \label{fig:KC_AC_all}
\end{figure}

Having explored the localization dynamics in the K-basis previous section, we will examine the time evolution of K-complexity and Arnoldi coefficients for various localization types discussed earlier in this section. We will also investigate whether the long-time behaviour of K-complexity and the variance of the Arnoldi coefficients can serve as reliable indicators for distinguishing between classical and quantum origins of localization in the system.\\

Fig \ref{fig:KC_AC_all} presents the time evolution of K-complexity (left panel) and Arnoldi coefficients (right panel) across different localization regimes. Panels (a) and (b) show the AR case over a short time scale ($t = 20$), where K-complexity exhibits perfect periodic oscillations with a period of two, indicating dynamics confined to two K-basis vectors. Consistently, the Arnoldi coefficients decay to zero after two steps. Panels (c) and (d) depict the CIL regime, where $\mathcal{C}$ shows a slight increase followed by persistent oscillations, reflecting limited wavefunction spread in the K-basis. The Arnoldi coefficients in this regime also display strong oscillations, highlighting the dominance of a few K-basis vectors in this strongly localized phase.\\

Fig \ref{fig:KC_AC_all}(e, f) illustrates the DL regime, where K-complexity exhibits linear growth for an extended period before eventually saturating, indicating a broad spread in the K-basis. The initial linear growth persists beyond the break time $t_b$ (marked by a dotted line), after which wavefunction growth is suppressed on the momentum basis. However, $\mathcal{C}$ saturates much later, as new K-basis vectors continue to form even after the onset of dynamical localization. Comparing panels (d) and (f), the Arnoldi coefficients in the DL regime show smaller amplitude oscillations compared to the CIL regime. This suggests that the variance of Arnoldi coefficients could serve as a useful indicator to distinguish between regularity and chaos in the system, despite both cases exhibiting localized quantum dynamics. Finally, Fig \ref{fig:KC_AC_all}(g, h) illustrates the PL regime, where K-complexity exhibits power-law growth over an extended period, indicating faster wavefunction spreading. The Arnoldi coefficients initially show a slight decay and revival, followed by a gradual decrease to zero, consistent with weaker localization compared to the other regimes. K-complexity with random deloclised vectors has been studied in the appendix \ref{app:random_state}\\

\section{Probes for the onset of chaos and Nature of Localization}
\label{Sec:6}

So far, we have thoroughly explored the extent of localization in the K-basis. Now, we shift our focus to understanding the nature of localization in quantum systems and ask whether these quantities can also capture the onset of chaos, even in the presence of quantum localization. In quantum systems, the transition from integrable to chaotic dynamics is often accompanied by significant changes in various physical quantities. To investigate whether this transition can be effectively detected, we analyze the average K-complexity ($\overline{\mathcal{C}}$), averaged over both time and initial states, as it has been shown to be sensitive to the onset of chaos \cite{rabinovici2022krylov}. Additionally, we explore the variance of the Arnoldi coefficients, $\sigma^2(h_{n,n-1})$, as another potential marker of chaos.

\begin{figure}[!t]
    \centering
    \includegraphics[width=0.8\linewidth]{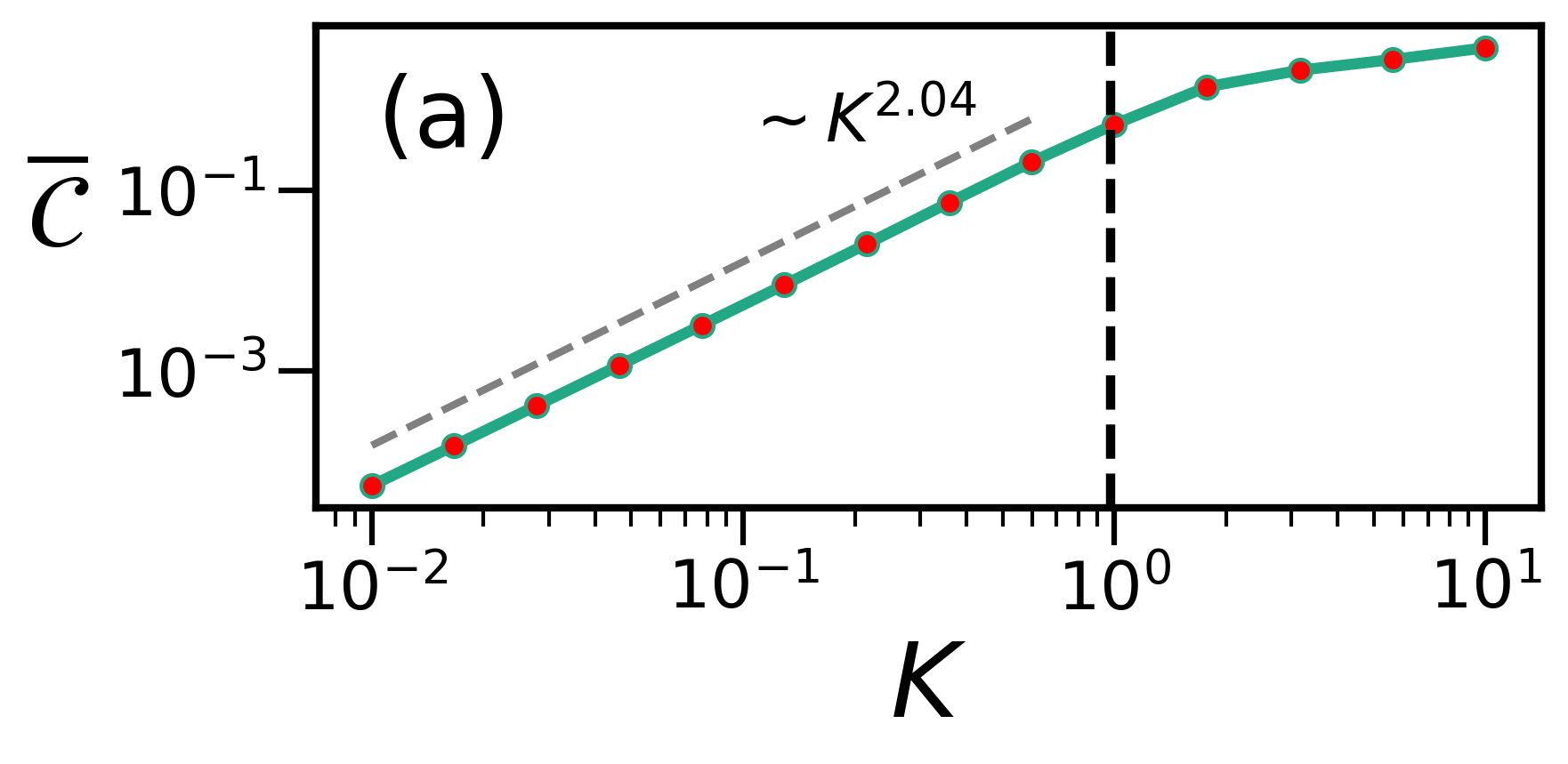}
    \includegraphics[width=0.8\linewidth]{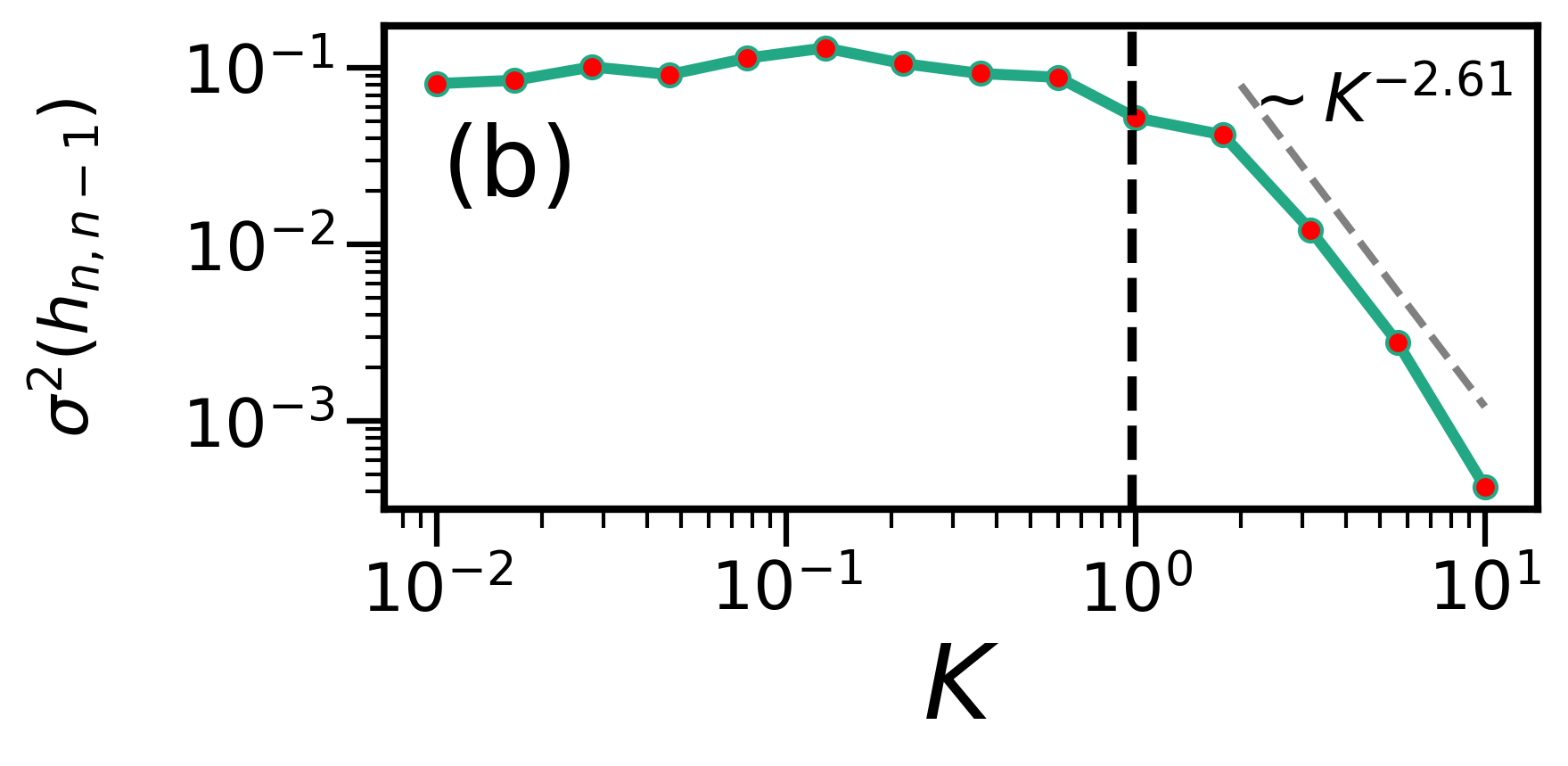}
    \caption{Comparison of average K-complexity ($\overline{\mathcal{C}}$) and the variance of Arnoldi coefficients $\sigma^{2}(h_{n,n-1})(t)$ as indicators of the integrability-to-chaos transition. (a) shows the growth of average K-complexity with respect to $K$ and (b) displays the variance of Arnoldi coefficients with respect to $K$.}
    \label{fig:avg_KC_AC_vs_K}
\end{figure}

In Fig.~\ref{fig:avg_KC_AC_vs_K}(a), the average K-complexity is plotted as a function of the kick strength $K$, it shows a power law increase with increasing $K$, with a slight change in slope near the onset of classical chaos. This transition closely corresponds to the breaking of the last invariant tori in the classical phase space of the kicked rotor, occurring at $K \approx 0.971$ \cite{santhanam2022quantum}. Although there is no sharp change in average K-complexity, there is a slight change in the slope near the transition point. Similarly, Fig.~\ref{fig:avg_KC_AC_vs_K}(b) shows the variance of Arnoldi coefficients, $\sigma^2(h_{n,n-1})$, as a function of $K$. It remains nearly constant for $K < 1$, whereas, for $K > 1$, it exhibits an approximate power-law decay, signalling a transition to chaotic dynamics. These results demonstrate that while K-complexity does not reliably indicate the onset of chaos and the variance of Arnoldi coefficients serves as a more effective diagnostic \cite{espanol2023assessing, pg2025dependence}.\\

When the system parameters are close to an integrable or regular regime, the dynamics are highly constrained, leading to significant oscillations in the Arnoldi coefficients. This occurs because only a few K-basis vectors are sufficient to capture the localized nature of the dynamics, and the overlap between consecutive basis vectors fluctuates strongly. In contrast, when the system enters a chaotic regime, the dynamics become more delocalized in the K-basis, with the wave function spreading across a larger number of basis vectors. As a result, the Arnoldi coefficients stabilize, exhibiting much smaller oscillations and values closer to one. This difference highlights the sensitivity of Arnoldi coefficients to the degree of localization and can also serve as an effective probe for distinguishing between integrable and chaotic behaviours in the system.\\

We now explore the dependence of K-complexity on the effective Planck's constant $\hbar_s$. The kicked rotor model, with its well-defined classical limit, provides an ideal framework to study how wavefunction spreading in the K-basis evolves during the quantum-to-classical transition as $\hbar_s$ varies. As $\hbar_s \to 0$, the system approaches the semiclassical regime, where quantum dynamics closely mimics classical behaviour for an extended period. Conversely, increasing $\hbar_s$ moves the system away from the semiclassical limit, causing quantum effects to emerge much earlier in the dynamics.\\

\begin{figure}[h!]
    \centering
    \includegraphics[width=0.49\linewidth]{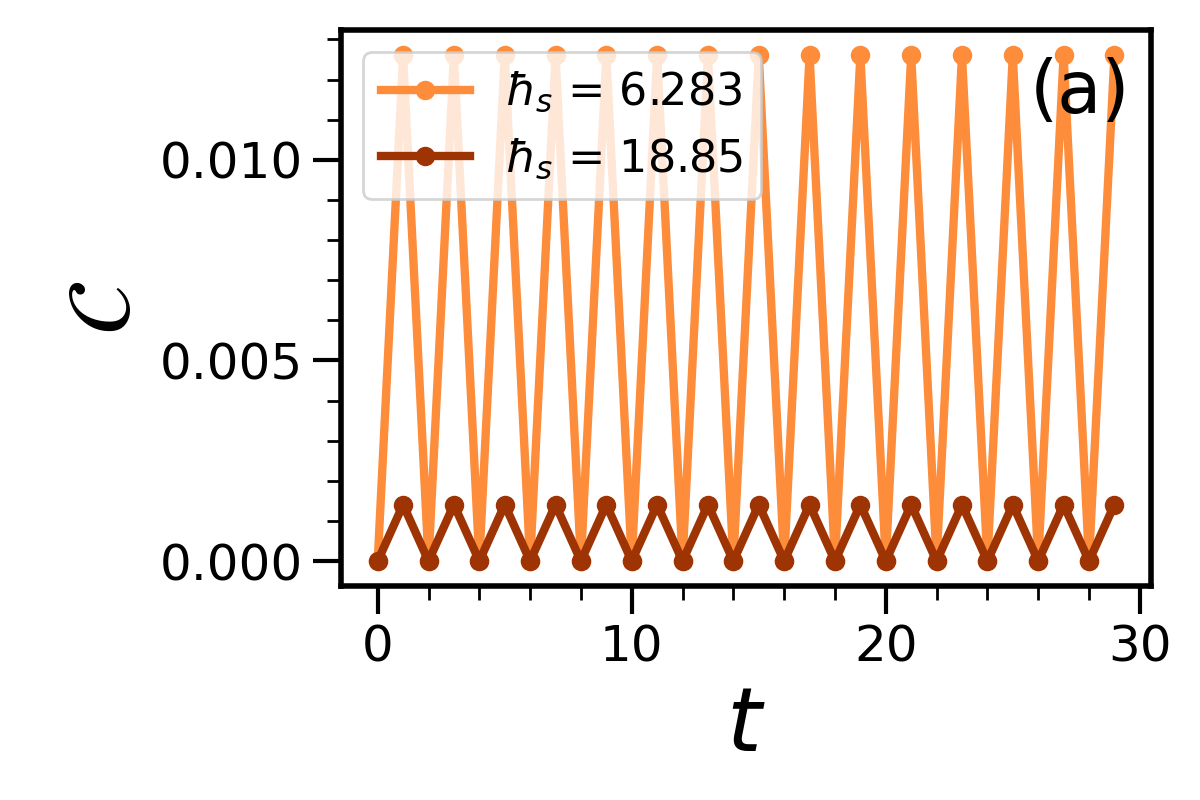}
    \includegraphics[width=0.49\linewidth]{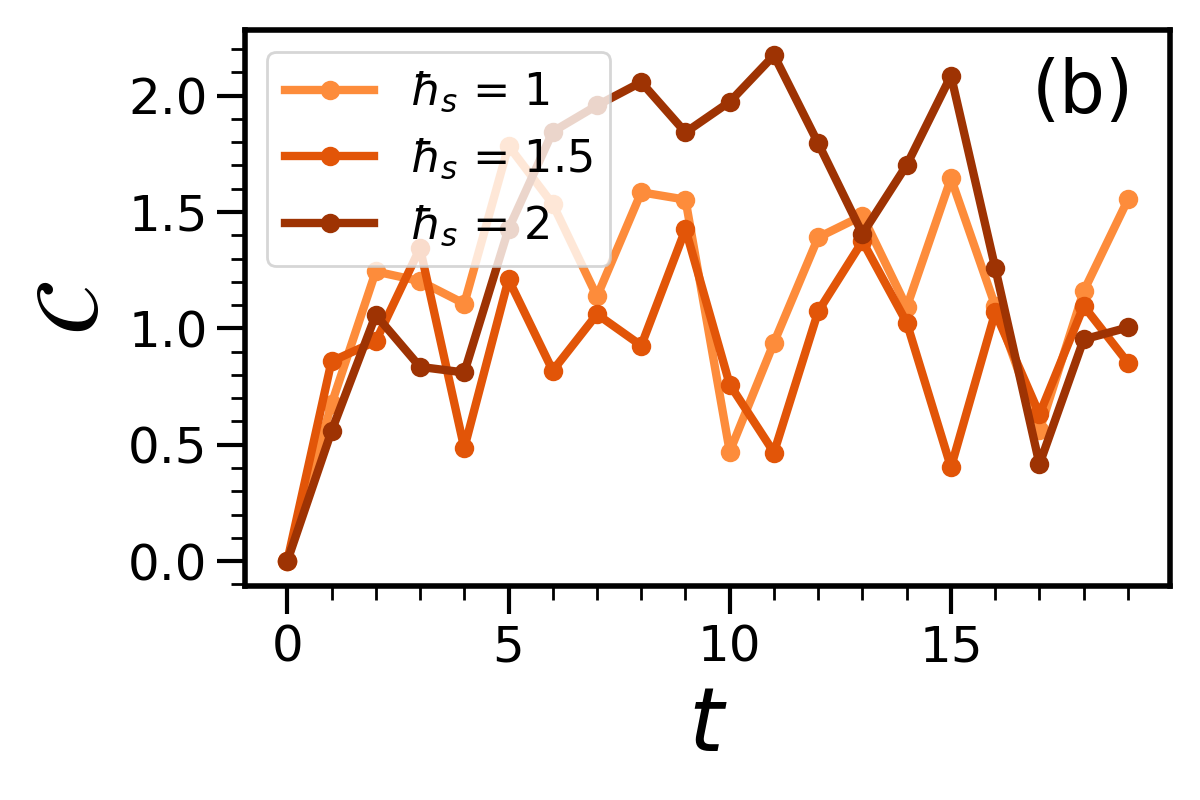}
    \includegraphics[width=0.49\linewidth]{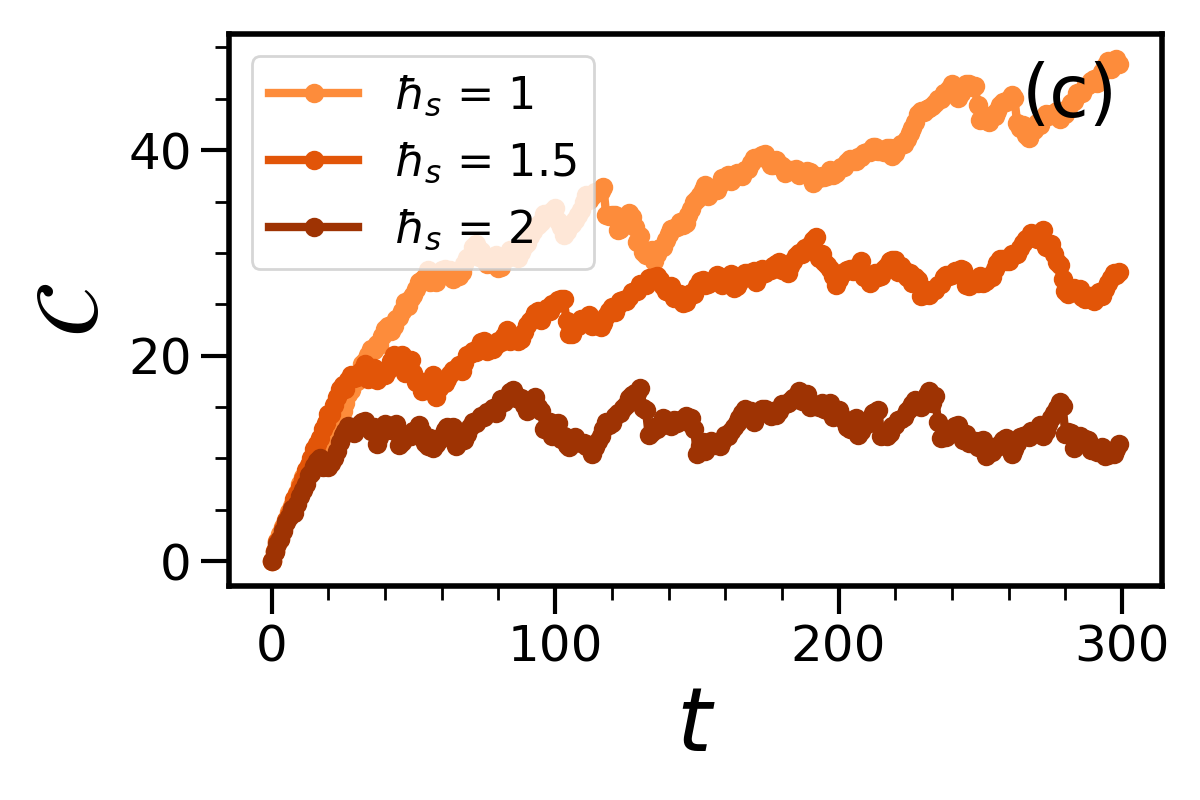}
    \includegraphics[width=0.49\linewidth]{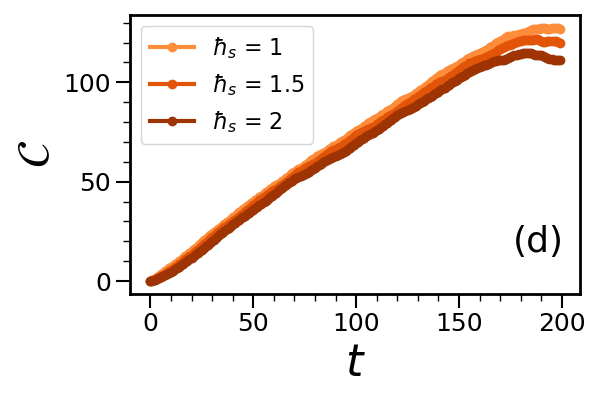}
    \caption{K-complexity ($\mathcal{C}$) as a function of the number of kicks ($t$) for different localization regimes with varying effective Planck's constants $\hbar_s$. (a) AR with $\hbar_s = (2\pi, 6\pi)$, (b) CIL, (c) DL, and (d) PL, each for $\hbar_s = (1.0, 1.5, 2.0)$.}
    \label{fig:KC_vs_time}
\end{figure}

We will first investigate how variations in $\hbar_s$ influence K-complexity across the four types of localization discussed earlier. Notably, for AR, $\hbar_s$ is restricted to discrete values of the form $\hbar_s = (4n + 2)$, where $n \in \mathbb{Z}$, and hence $\hbar_s$ cannot be varied continuously. For the other three localization types, $\hbar_s$ can be varied continuously. Fig. \ref{fig:KC_vs_time}(a) illustrates the case of AR for two values of $\hbar_s = (2\pi, 6\pi)$. As $\hbar_s$ increases, the wave packet undergoes stronger localization, resulting in a noticeable reduction in K-complexity. Nevertheless, periodic oscillations persist, and the dynamics remain confined to just two K-basis vectors, reflecting the minimal complexity of the AR regime.\\

Fig \ref{fig:KC_vs_time}(b) depicts the CIL regime for $\hbar_s = (1.0, 1.5, 2.0)$, where $\mathcal{C}$ exhibits oscillatory behavior over time. Notably, no clear trend emerges with varying $\hbar_s$, as the CIL regime is strongly influenced by regular structures in the classical phase space. Consequently, changes in $\hbar_s$ have a non-monotonic impact on its behaviour. Fig \ref{fig:KC_vs_time}(c) explores the DL regime for $\hbar_s = (1.0, 1.5, 2.0)$, where $\mathcal{C}$ exhibits noticeable variations in saturation values with changing $\hbar_s$. As $\hbar_s$ increases, the saturation value of $\mathcal{C}$ decreases, reflecting the inverse relationship with the localization length ($\xi$) in the momentum basis  $\xi \sim \hbar_s^{-2}$. Since K-complexity measures wavefunction spread in the K-basis rather than the momentum basis, hence we anticipate a similar power-law dependence with $\hbar_s$. Finally, Fig.~\ref{fig:KC_vs_time}(d) depicts the PL regime for $\hbar_s = (1.0, 1.5, 2.0)$, where in the long time limit $\mathcal{C}$ exhibits a noticeable dependence on $\hbar_s$. A clear distinction emerges with $\mathcal{C}$ consistently decreasing as $\hbar_s$ increases.\\

\begin{figure}[h!] 
\centering 
\includegraphics[width=0.49\linewidth]{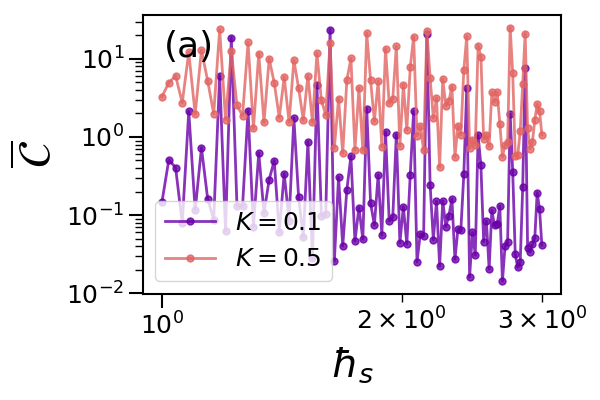} 
\includegraphics[width=0.49\linewidth]{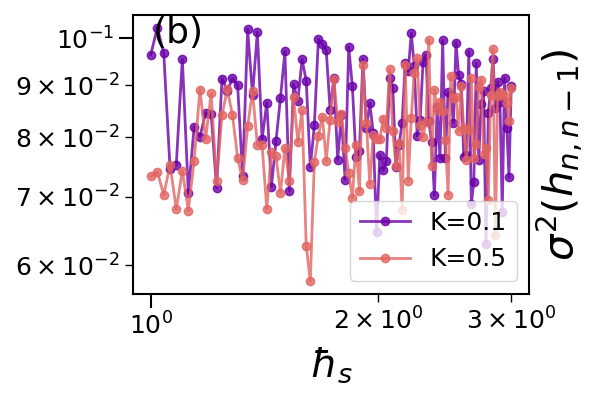} 
\includegraphics[width=0.49\linewidth]{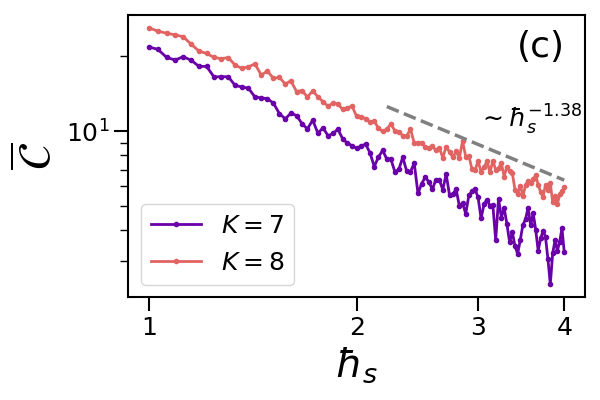} 
\includegraphics[width=0.49\linewidth]{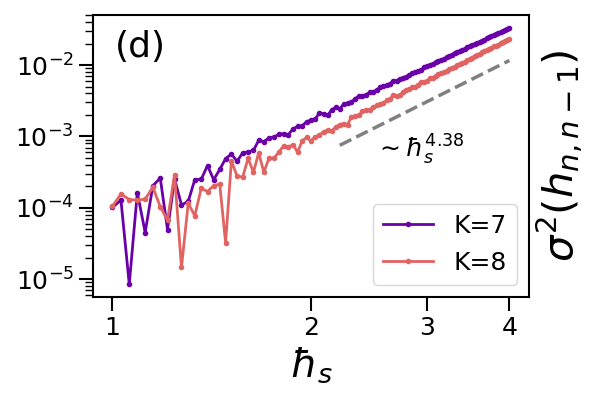} 
\caption{Scaling of average K-complexity ($\overline{\mathcal{C}}$) and variance of Arnoldi coefficients ($\sigma^{2}(h_{n,n-1})$) with $\hbar_s$ in different localization regimes. (a) $\overline{\mathcal{C}}$ vs $\hbar_s$ in the CIL regime and (b) $\sigma^{2}(h_{n,n-1})$ vs $\hbar_s$ in the CIL regime, both showing strong oscillations without clear scaling. (c) $\overline{\mathcal{C}}$ vs. $\hbar_s$ in the DL regime, exhibiting a power-law decay. (d) $\sigma^{2}(h_{n,n-1})$ vs. $\hbar_s$ in the DL regime, displaying a power-law increase.}
\label{fig:KC_AC_scaling} 
\end{figure}

Now we will study the scaling of K-complexity and the variance of Arnoldi coefficients (both averaged over time and initial states) as functions of $\hbar_s$. Fig. \ref{fig:KC_AC_scaling}(a) shows  $\overline{\mathcal{C}}$, while Fig.~\ref{fig:KC_AC_scaling}(b) depicts $\sigma^{2}(h_{n,n-1})$, as a function of $\hbar_s$ with $K$ chosen to be in the CIL regime. Interestingly, neither $\overline{\mathcal{C}}$ nor $\sigma^{2}(h_{n,n-1})$ exhibits a clear scaling behavior with $\hbar_s$. Instead, both quantities show pronounced oscillations as $\hbar_s$ is varied. This lack of a distinct scaling with $\hbar_s$ is characteristic of the CIL regime, where the evolution of the wavefunction is constrained by the persistence of regular structures in phase space. As a result, the system remains in a near-integrable regime, leading to a non-trivial interplay between the localization of the wavepacket and variations in $\hbar_s$. In contrast, the case of DL shows a clear scaling of the saturation value of K-complexity, which decreases with a power law fashion, as shown in Fig.~\ref{fig:KC_AC_scaling}(c). Similarly, the variance of the Arnoldi coefficients increases with $\hbar_s$, as depicted in Fig.~\ref{fig:KC_AC_scaling}(d).

\section{Conclusions}
\label{Sec:7}
Krylov complexity (K-complexity) is an emerging and popular measure of operator and state complexity. While quantum chaotic systems and the associated ergodic states have been studied before in the Krylov basis, the other extreme of localized states are poorly understood in the Krylov basis. How does quantum localization -- a quantum phenomenon of interest across condensed matter physics and quantum chaos -- manifest in the Krylov basis ? We answer this question by investigating localization dynamics arising in two variants of the QKR models; the standard kicked rotor with a smooth kicking potential and a kicked rotor with a non-smooth potential. We focus on the wavefunction spreading in the Krylov basis and the evolution of K-complexity for four distinct localization scenarios -- (i) localization induced by classically invariant structures, (ii) localization due to quantum anti-resonance, (iii) dynamical localization arising from quantum interference effects, and (iv) a weaker form of power-law localization.

Firstly, we show that the degree of localization in the K-basis is reflected in the corresponding probability amplitudes and inverse participation ratios, while K-complexity exhibited distinct growth patterns depending on the strength of localization. Secondly, we show that, in the K- basis, classically induced localization effects can be distinguished from quantum localization effects. The long-time averaged K-complexity and fluctuations in Arnoldi coefficients reveal that when localization arises from regular structures in the classical phase space, no clear scaling with effective Plancks constant $\hbar_s$ is observed. However, for localization due to quantum effects and even if not associated with regular classical structures, a systematic scaling with $\hbar_s$ emerges. This work highlights how classical phase space structures and quantum interference shape localization dynamics in the K-basis. It also shows that K-complexity effectively classifies different types of localization and distinguishes their origins -- whether classical or quantum.\\

Interestingly, in the quantum kicked rotor models, even if quantum ergodicity is absent due to the presence of quantum localization, the variance of Arnoldi coefficients effectively capture the onset of classical chaos. The latter is more sensitive measure to probe the integrability to chaos transition than K-complexity. This work can be extended to study localization in the presence of interactions; this includes dynamical many-body localization, delocalization and prethermal effects in the K-basis \cite{nambudiripad2024chaos, paul2024faster, sadia2022prethermalization}. Recent works have shown that K-complexity has a pronounced dependence on the choice of the initial operator or state, raising doubts about its reliability as a definitive chaos quantifier \cite{pg2025dependence}. In particular, K-complexity has monotonic dependence on the spread of the initial state in the eigenbasis implying that changes in initial conditions can significantly alter saturation values of K-complexity even in the chaotic regime. Our analysis further shows that when a randomly delocalized vector is chosen as the initial state, it does not appear feasible to distinguish between various degrees of localization. If K-complexity must be regarded as a reliable measure of chaos, a more effective strategy for initial state seeding must be adopted \cite{craps2025multiseed}.

\appendix

\section{Spectral Statistics}
\label{app:sprectral_stats}
Given a discrete eigenvalue spectrum $\lambda_1, \lambda_2, \dots \lambda_N$, spacing is defined as $s = \lambda_{n+1}-\lambda_n$. The symmetry-reduced level statistics of classically integrable systems show the Poissonian distribution of spacings, indicating no level of repulsion, as per the Berry-Tabor conjecture \cite{berry1977level}. In contrast, the Bohigas-Giannoni-Schmit conjecture suggests that chaotic systems follow Wigner-Dyson statistics, characterized by level repulsion, signalling quantum chaos \cite{bohigas1984characterization}. While spectral statistics reveal whether a system is chaotic or integrable, they don’t fully capture the degree of localization in quantum systems. Spectral statistics focus on energy level correlations but do not directly probe the spatial or dynamical aspects of localization. In this section, we examine spectral correlations for varying degrees of localization in the kicked rotor model using the quasi-energies of the Floquet operator.\\

The relationship between the Floquet operator $U_F$ and its eigenstate $\ket{\psi_\varphi}$ is expressed as:
\begin{align}
    U_F \ket{\psi_\varphi} = e^{i\varphi} \ket{\psi_\varphi},
\end{align}
where $\varphi$ represents the quasienergy associated with the eigenstate $\ket{\psi_\varphi}$. The spacing between consecutive eigenvalues is given by $s_n = \frac{N}{2\pi} (\varphi_n - \varphi_{n-1})$, where $N$ is the total number of eigenvalues, and $s_n$ represents the spacing between the $n$-th and $(n-1)$-th eigenvalue. To quantify the distribution of these spacings, we define the following normalized parameter $\langle \tilde{r} \rangle$:

\begin{align}
    \langle \tilde{r} \rangle = \frac{1}{D}\sum_{n=1}^D \frac{\min(s_n, s_{n-1})}{\max(s_n, s_{n-1})},
\end{align}
where $D$ is the total number of eigenvalues and $\langle \tilde{r} \rangle$ represents the average ratio of consecutive level spacings. 
For integrable systems, where the eigenvalues are expected to be spaced in a Poissonian fashion, the value is $\langle \tilde{r} \rangle_P \approx 0.38629$ and for chaotic systems, modelled by the GOE ensemble in RMT, the value is $\langle \tilde{r} \rangle_{GOE} \approx 0.53590.$ \cite{atas2013distribution}\\
\begin{figure}[h!]
    \centering
    \includegraphics[width=1.0\linewidth]{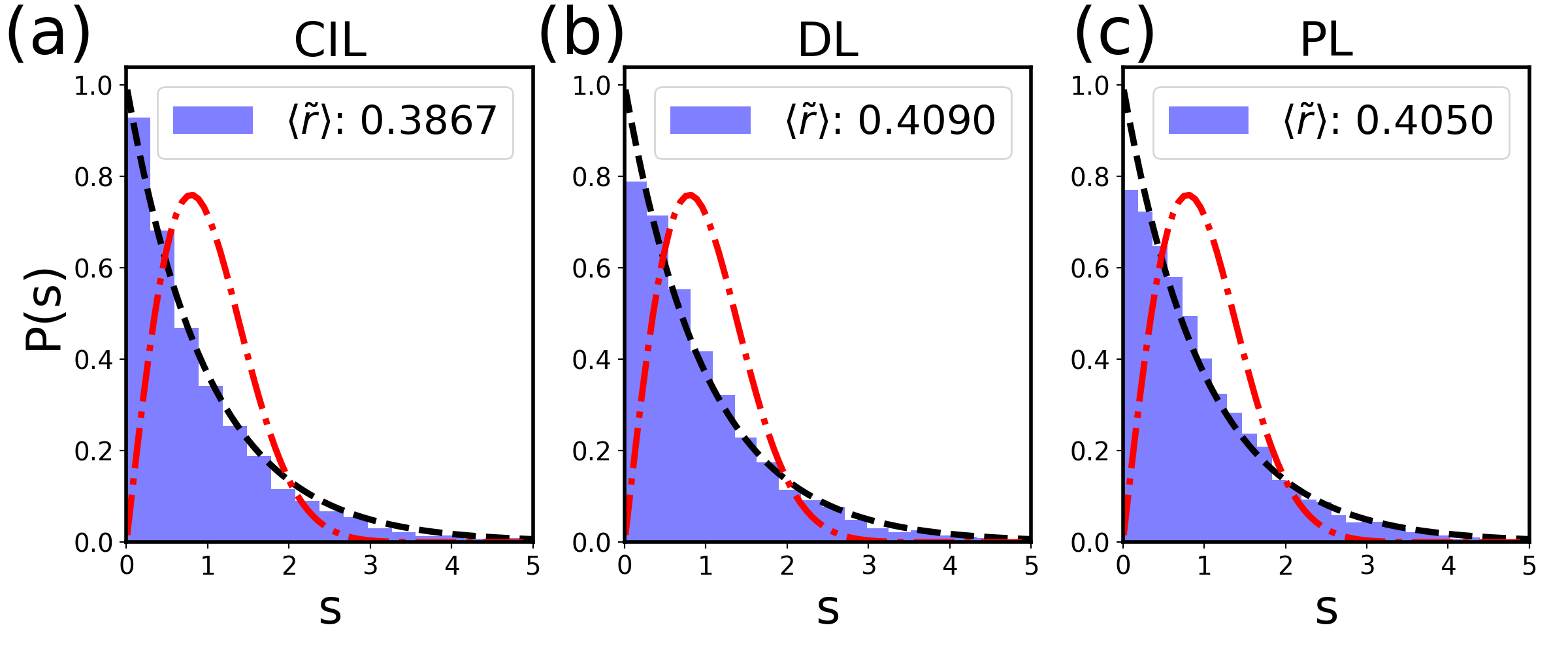}
    \caption{Distribution of level spacings $s$ for diff regimes of localization. The histograms are compared with the Poisson and Gaussian Orthogonal Ensemble (GOE) distributions, represented by dashed and dash-dotted lines. The plots correspond to the regimes (a) CIL, (b) DL, and (c) PL.}
    \label{fig:level_stats}
\end{figure}
The level spacing distributions $P(s)$ for different localization regimes are shown in Fig.(\ref{fig:level_stats}), with histograms representing numerical results and dashed/dash-dotted lines corresponding to Poisson and GOE statistics. In all three regimes—CIL (Fig. \ref{fig:level_stats}(a)), DL (Fig. \ref{fig:level_stats}(b)), and PL (Fig. \ref{fig:level_stats}(c))—the distribution consistently follows Poisson statistics. This indicates that level spacing is not a suitable measure to study the extent or type of localization. Instead, $P(s)$ serves as a robust diagnostic tool to distinguish localization from chaos, as Poisson statistics are characteristic of localized systems.

\section{System size dependence}
\label{app:system_size}
\begin{figure}[H]
\centering 
\includegraphics[width=0.6\linewidth]{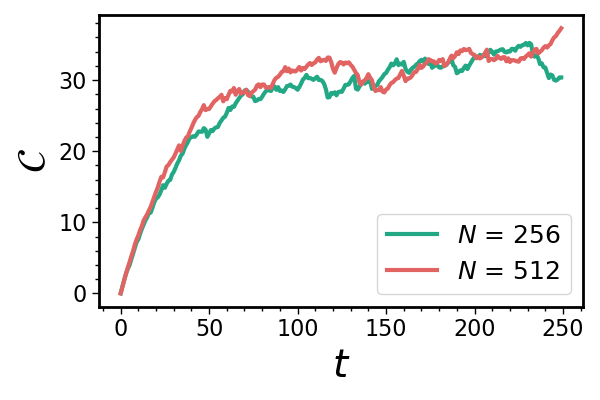} 
\caption{Dependence of K-complexity saturation values on system size $N$ under dynamical localization. The figure demonstrates that even as $N$ increases, the saturation values remain nearly constant, indicating that the observed saturation is primarily due to dynamical localization rather than the finite size of the K-basis.} 
\label{fig:KC_DL_vary_N} 
\end{figure}

In this section, we analyze the dependence of K-complexity saturation values on the system size $N$ across different dynamical regimes. By studying the evolution of K-complexity for varying $N$, we investigate whether the observed saturation is an artefact of the finite size of the K-basis or a genuine feature of the underlying dynamics. The dependence of K-complexity saturation values on system size $N$ is illustrated in Fig.~\ref{fig:KC_DL_vary_N} for the case of DL. The plots show that the saturation values of K-complexity remain nearly constant across different system sizes, even as $N$ increases. This invariance indicates that saturation is a genuine feature of the underlying dynamics, primarily attributed to dynamical localization rather than an artefact of the finite size of the K-basis.

\section{K-Complexity for Random Initial States}
\label{app:random_state}
In this section, we investigate the behaviour of K-complexity for random delocalized initial states in momentum space. These states are generated by applying a Haar random unitary transformation to a delta-localized state. Using this random state, we construct the K-basis and compute the corresponding K-complexity. In the AR regime, the behaviour of K-complexity remains unaffected, continuing to exhibit perfect oscillations. However, in the CIL regime, K-complexity behaves distinctly differently from the initially localized state. It grows and saturates, much like in the DL case, failing to capture the influence of regular structures in phase space, as shown in Fig. (\ref{fig:KC_vs_time})(b). Due to the delocalized nature of the random initial state, it becomes less effective in distinguishing between different types of localization. As a result, the K-complexity for random delocalized states is significantly larger than that of localized states, highlighting a strong initial state dependence, as recently emphasized in \cite{pg2025dependence}.In the PL regime, K-complexity for random initial states exhibits an initial linear growth followed by rapid saturation at long times. This behaviour contrasts sharply with that of delta-localized initial states, where saturation occurs at a much later time step, as illustrated in Fig. (\ref{fig:KC_vs_time})(d). These clear distinctions underscore the sensitivity of K-complexity to the choice of initial states.

\begin{figure}[H]
    \centering
    \includegraphics[width=0.49\linewidth]{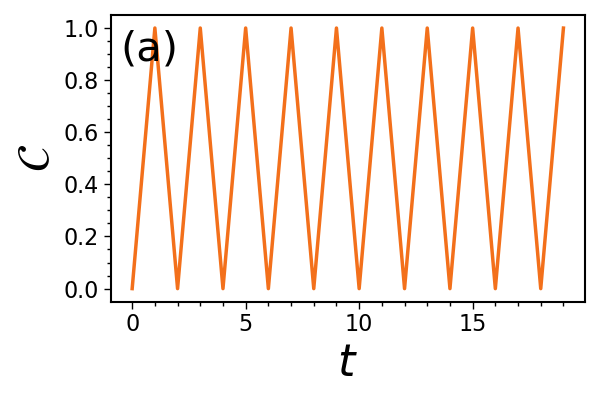}
    \includegraphics[width=0.49\linewidth]{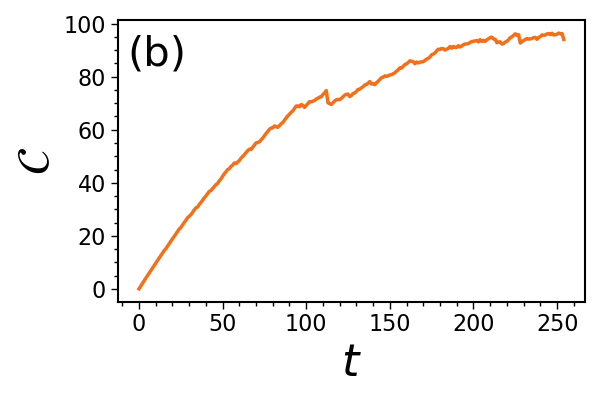}
    \includegraphics[width=0.49\linewidth]{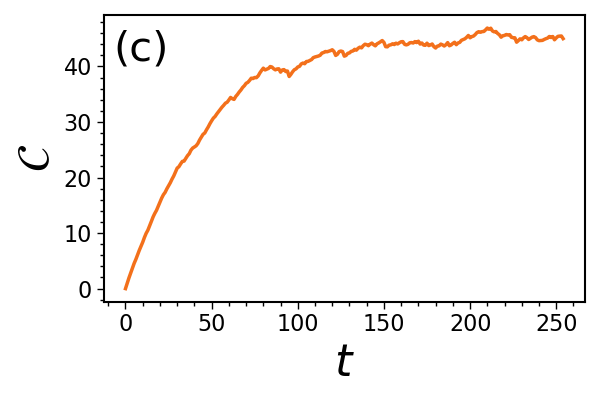}
    \includegraphics[width=0.49\linewidth]{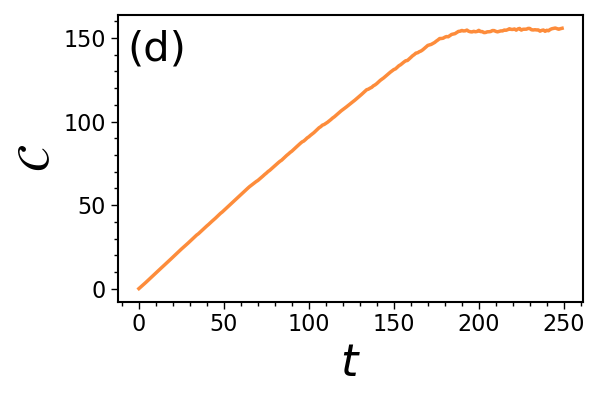}
    \caption{K-complexity ($\mathcal{C}$) as a function of the number of kicks $t$, for four distinct regimes: (a) AR regime,  (b) CIL regime,  (c) DL regime, and  (d) PL regime.}
    \label{fig:random_KC}
\end{figure}

\bibliography{ref}
\bibliographystyle{elsarticle-num}

\end{document}